# BINARY TREES AND NUMBER OF STATES IN BUDDY SYSTEMS

ZOLTÁN KÁSA and LEON TÂMBULEA

Babeş–Bolyai University, Faculty of Mathematics
Str. Kogălniceanu 1, 3400 Cluj-Napoca, Romania

**Abstract.** In this paper we give a formula for the number of leaves in the set of the binary trees with $n$ nodes, and one for the number of binary trees with $n$ nodes, which have $k$ leaves. The latter is used to compute the number of states in buddy systems. In the last section we prove three combinatorial identities using the formulae given in the first section.

## 1. Binary trees

Let $b_n^{(k)}$ denote the number of binary trees with $n$ nodes and $k$ leaves (terminal nodes). For $k > [(n+1)/2]$ we have $b_n^{(k)} = 0$. The number of leaves in the set of the binary trees with $n$ nodes will be denoted by $b_n$.

In this section we prove the following two theorems:

**Theorem 1.** *The number $b_n^{(k)}$ of the binary trees with $n$ nodes and $k$ leaves is given by the formula:*

$$(1) \qquad b_n^{(k)} = \frac{1}{n}\binom{n}{2k-1}\binom{2k}{k}2^{n-2k}.$$

**Theorem 2.** *The number $b_n$ of leaves in the set of the binary trees with $n$ nodes is given by:*

$$(2) \qquad b_n = \binom{2n-2}{n-1}.$$

**Proof of the Theorem 1.** For $b_n^{(k)}$ we have the following recurrence equation:

$$(3) \qquad b_n^{(k)} = 2b_{n-1}^{(k)} + \sum_{i=1}^{n-2}\sum_{j=1}^{k-1} b_i^{(j)} b_{n-i-1}^{(k-j)}.$$

For small values of $k$ we easily obtain the appropriate formulae for $b_n^{(k)}$, using (3). E.g. $b_n^{(1)} = 2^{n-1}$, $n \geq 1$; $b_n^{(2)} = \binom{n-1}{2}2^{n-3}$, $n \geq 3$; $b_n^{(3)} = \binom{n-1}{4}2^{n-4}$,

$n \geq 5$ etc. By generalization we obtain:

(4) $$b_n^{(k)} = \binom{n-1}{2k-2} c_k 2^{n-3\left[\frac{k}{2}\right]-s}$$

where $s = 1$ for $k$ even, $s = 2$ for $k$ odd, and $c_k$ are coefficients, which will be calculated later. The proof of (4) can be made by induction on $k$. For this we shall use also the formula 3.3 from [1] in the following form:

$$\sum_{i=r}^{n-3-s} \binom{i}{r}\binom{n-3-i}{s} = \binom{n-2}{r+s+1}.$$

Let us suppose (4) true for values less than $k$. If $k = 4m$ we have from (3), eliminating the zero terms:

$$b_n^{(4m)} = 2b_{n-1}^{(4m)} + 2\{b_1^{(1)} b_{n-2}^{(4m-1)} + \ldots + b_{n-8m+2}^{(1)} b_{8m-3}^{(4m-1)} +$$
$$+ b_3^{(2)} b_{n-4}^{(4m-2)} + \ldots + b_{n-8m+4}^{(2)} b_{8m-5}^{(4m-2)} + \ldots + b_{4m-3}^{(2m-1)} b_{n-4m+2}^{(2m+1)} + \ldots +$$
$$+ b_{n-4m-2}^{(2m-1)} b_{4m+1}^{(2m+1)}\} + b_{4m-1}^{(2m)} b_{n-4m}^{(2m)} + \ldots + b_{n-4m}^{(2m)} b_{4m-1}^{(2m)} =$$
$$= 2b_{n-1}^{(4m)} + 2\left\{c_1 c_{4m-1} 2^{n-6m-2} \cdot \sum_{k=0}^{n-8m+1} \binom{k}{0}\binom{n-3-k}{8m-4} +\right.$$
$$+ c_2 c_{4m-2} 2^{n-6m-3} \cdot \sum_{k=2}^{n-8m+3} \binom{k}{2}\binom{n-3-k}{8m-6} +$$
$$+ \ldots + c_{2m-1} c_{2m+1} 2^{n-6m-2} \cdot \sum_{k=4m-4}^{n-4m-3} \binom{k}{4m-4}\binom{n-3-k}{4m}\right\} +$$
$$+ c_{2m} c_{2m} 2^{n-6m-3} \cdot \sum_{k=4m-2}^{n-4m-1} \binom{k}{4m-2}\binom{n-3-k}{4m-2} =$$
$$= 2b_{n-1}^{(4m)} + 2^{n-6m-1} \cdot \binom{n-2}{8m-3} \cdot c_{4m}.$$

From this we obtain

$$b_n^{(4m)} = \binom{n-1}{8m-2} \cdot c_{4m} \cdot 2^{n-6m-1}$$

which has the form of (4).

The coefficients satisfy the following equation:

(5) $$c_{4m} = c_1 c_{4m-1} + \frac{1}{2} c_2 c_{4m-2} + \ldots + \frac{1}{2} c_{2m-2} c_{2m+2} + c_{2m-1} c_{2m+1} + \frac{1}{4} c_{2m}^2.$$

In the same way we find the following:

$$b_n^{(4m+1)} = \binom{n-1}{8m} c_{4m+1} 2^{n-6m-2},$$

(6) $\quad c_{4m+1} = \dfrac{1}{2}\Big(c_1 c_{4m} + c_2 c_{4m-1} + \ldots + c_{2m-1} c_{2m+2} + c_{2m} c_{2m+1}\Big),$

$$b_n^{(4m+2)} = \binom{n-1}{8m+2} c_{4m+2} 2^{n-6m-4},$$

(7) $\quad c_{4m+2} = c_1 c_{4m+1} + \dfrac{1}{2} c_2 c_{4m} + \ldots + c_{2m-1} c_{2m+3} + \dfrac{1}{2} c_{2m} c_{2m+2} + \dfrac{1}{2} c_{2m+1}^2,$

$$b_n^{(4m+3)} = \binom{n-1}{8m+4} c_{4m+3} 2^{n-6m-5},$$

(8) $\quad c_{4m+3} = \dfrac{1}{2}\Big(c_1 c_{4m+2} + c_2 c_{4m+1} + \ldots + c_{2m+1} c_{2m+2}\Big).$

To find the value of $c_k$ for $k \geq 1$, we put

(9) $$d_{2k} = \sqrt{2} \cdot c_{2k}$$
$$d_{2k+1} = c_{2k+1}$$

and from (5)–(8) we have

$$d_{2k} = \dfrac{1}{\sqrt{2}}\Big(d_1 d_{2k-1} + d_2 d_{2k-2} + \ldots + d_{k-1} d_{k+1} + \dfrac{1}{2} d_k^2\Big),$$

$$d_{2k+1} = \dfrac{1}{\sqrt{2}}(d_1 d_{2k} + d_2 d_{2k-1} + \ldots + d_k d_{k+1}).$$

Consider now the following generating function

$$D(z) = \sum_{i \geq 1} d_i z^i.$$

This function satisfies the following equation:

$$D^2(z) = 2\sqrt{2}(D(z) - 2z).$$

The solution which meets our requirements is

$$D(z) = \sqrt{2} + \sqrt{2 - 4\sqrt{2}z}.$$

By expansion we have

$$D(z) = \sqrt{2} - \sqrt{2} \sum_{i \geq 0} \binom{1/2}{i}(-1)^i (2\sqrt{2}\, z)^i =$$

$$= \sqrt{2} - \sqrt{2} \sum_{i \geq 0} \dfrac{(-1)^{i-1}}{2^{2i}(2i-1)} \binom{2i}{i}(-1)^i (2\sqrt{2})^i \cdot z^i = \sum_{i \geq 1} \binom{2i}{i} \dfrac{1}{(\sqrt{2})^{i-1}(2i-1)} z^i$$

and

(10) $$d_i = \binom{2i}{i} \frac{1}{(\sqrt{2})^{i-1}(2i-1)}.$$

Therefore, from (4), (9) and (10) it follows

$$b_n^{(k)} = \binom{n-1}{2k-2}\binom{2k}{k}\frac{1}{2k-1} 2^{n-2k} = \frac{1}{n}\binom{n}{2k-1}\binom{2k}{k} \cdot 2^{n-2k}. \quad \square$$

**Proof of the Theorem 2.** The number of binary trees with $n$ nodes is

$$a_n = \frac{1}{n+1}\binom{2n}{n}.$$

There are $a_k a_{n-k-1}$ binary trees with $n$ nodes, which have a left subtree with $k$ nodes, and a right subtree with $n-k-1$ nodes. In such binary trees with a given left subtree with $k$ nodes we have $a_{n-k-1} \cdot v_1 + b_{n-k-1}$ leaves, where $v_1$ is the number of the leaves in the fixed left subtree. If we consider all the $a_k$ left subtrees, we have

$$\sum_{i=1}^{a_k}(a_{n-k-1}v_i + b_{n-k-1}) = a_{n-k-1}\cdot b_k + a_k \cdot b_{n-k-1}$$

leaves. By summing up these for $k = 0, 1, \ldots, n-1$, and taking into account that $b_0 = 0$ and $a_0 = 1$, we have

(11) $$b_n = 2(a_0 b_{n-1} + a_1 b_{n-2} + \ldots + a_{n-2} b_1).$$

Consider the following generating functions:

$$A(z) = \sum_{i\geq 0} a_i z^i, \qquad B(z) = \sum_{i\geq 1} b_i z^i.$$

These functions satisfy the equation

$$A(z)B(z) = \frac{1}{2z}B(z) - \frac{1}{2}.$$

From this we obtain

$$B(z) = \frac{z}{1 - 2zA(z)}.$$

In [2] it is proved that

$$A(z) = \frac{1}{2z}(1 - \sqrt{1-4z})$$

and we have

(12) $$B(z) = \frac{z}{\sqrt{1-4z}}.$$

Using the formula

$$\frac{1}{\sqrt{1-z}} = \sum_{n\geq 0}\binom{2n}{n}2^{-2n}z^n$$

(see [1]) we obtain

$$B(z) = z\sum_{n\geq 0}\binom{2n}{n}2^{-2n}(4z)^n = \sum_{n\geq 0}\binom{2n}{n}z^{n+1} = \sum_{n\geq 1}\binom{2n-2}{n-1}z^n.$$

Consequently

$$b_n = \binom{2n-2}{n-1}. \quad \Box$$

## 2. Number of states in buddy systems

A buddy system is a dynamic storage allocation method [2], in which storage is provided in prescribed size. In the simplest buddy system, called binary buddy system, if a free block of a particular size is not available to satisfy a request, a larger free block is split into two smaller blocks, which are called buddies. The splitting continues until an appropriate block is found, which is allocated, and the other is placed on the free space list. Two free blocks may be merged in a single block if and only if they are buddies. In the binary buddy system block sizes are $2^i$.

A state in a buddy system is a storage configuration, with allocated and free blocks. A $k$-state is a state with $k$ blocks.

To count the number of $k$-states in binary buddy system, we make a correspondence between the allocation with buddy system and binary trees.

A regular binary tree is a binary tree in which all internal nodes have exactly two children. The entire storage pool corresponds to the root, buddies resulting by a split correspond to children of a given node. The number $N(k+1)$ of $(k+1)$-states is given by the next

**Lemma.** *If* $k \geq 1$ *then*

$$N(k+1) = 2^{k+1}\sum_{i=1}^{s}\left(\frac{3}{4}\right)^{p_i},$$

*where* $s = \dfrac{1}{k+1}\dbinom{2k}{k}$ *is the number of regular binary trees with $k+1$ leaves (or the number of binary trees with $k$ nodes), and $p_i$ is the number of buddies in the state which corresponds to the $i^{th}$ regular binary tree.*

**Proof.** If $p_i$ is the number of buddies, let $n_i$ be the number of other blocks in the $i^{th}$ $(k+1)$-state. Then $2p_i + n_i = k+1$.
But

$$N(k+1) = \sum_{i=1}^{s} 2^{n_i}3^{p_i} = 2^{k+1}\sum_{i=1}^{s}\left(\frac{3}{4}\right)^{p_i}. \quad \Box$$

If we delete all leaves of a regular binary tree with $2k+1$ nodes, we obtain a binary tree with $k$ nodes. If we complete this binary tree with nodes to obtain the corresponding regular binary tree, we observe that all leaves in the binary tree produce pairs of leaves (i.e. buddies in state) in the regular binary tree.
Then

$$N(k+1) = 2^{k+1} \sum_{i=1}^{\left[\frac{k+1}{2}\right]} b_k^{(i)} \left(\frac{3}{4}\right)^i$$

and we have the following

**Theorem 3.** *The number of $(k+1)$-states in binary buddy system is given by*

$$N(k+1) = \frac{3 \cdot 2^{2k-2}}{k} \sum_{i=0}^{\left[\frac{k-1}{2}\right]} \binom{k}{i}\binom{k-i}{i+1} 2^{-4i} 3^i.$$

**Proof.**

$$N(k+1) = 2^{k+1} \sum_{i=1}^{\left[\frac{k+1}{2}\right]} b_k^{(i)} \left(\frac{3}{4}\right)^i = 2^{k+1} \sum_{i=1}^{\left[\frac{k+1}{2}\right]} \frac{1}{k}\binom{k}{2i-1}\binom{2i}{i} 2^{k-2i} \left(\frac{3}{4}\right)^i =$$

$$= \frac{1}{k} \cdot 2^{2k-3} \sum_{i=0}^{\left[\frac{k-1}{2}\right]} \binom{k}{2i+1}\binom{2i+2}{i+1} 2^{-4i} 3^{i+1} =$$

$$= \frac{3}{k} 2^{2k-2} \sum_{i=0}^{\left[\frac{k-1}{2}\right]} \binom{k}{2i+1}\binom{2i+1}{i} 2^{-4i} 3^i =$$

$$= \frac{3 \cdot 2^{2k-2}}{k} \sum_{i=0}^{\left[\frac{k-1}{2}\right]} \binom{k}{i}\binom{k-i}{i+1} 2^{-4i} 3^i. \quad \square$$

### 3. Combinatorial identities

In this section we give some combinatorial identities that follow from the results of Section 1.

From these, the formula given in Consequence 2 is identical to "3.92" in [1]. We give here a new proof for it. The formulae given in Consequences 1 and 3 seem to be new.

**Consequence 1.** If $n \geq 1$ then

$$\sum_{k=1}^{n} \frac{2^{2(n-k)}}{2k-1} \binom{2k}{k} = 2^{2n} - \binom{2n}{n}.$$

**Proof.** We give another expansion of $B(z)$ in Section 1. From (12)

$$B(z) = \frac{z}{\sqrt{1-4z}} = \frac{z\sqrt{1-4z}}{1-4z},$$

but

$$\frac{z}{1-4z} = z \sum_{n\geq 0} (4z)^n,$$

and

$$\sqrt{1-4z} = \sum_{n\geq 0} \binom{1/2}{n}(-4z)^n = \sum_{n\geq 0} \frac{(-1)^{n-1}}{2^{2n}(2n-1)}\binom{2n}{n}(-4z)^n =$$

$$= \sum_{n\geq 0} \frac{-1}{2n-1}\binom{2n}{n}z^n.$$

Therefore

$$B(z) = \left(\sum_{n\geq 0} 4^{n-1} z^n\right)\left\{\sum_{n\geq 0} \frac{-1}{2n-1}\binom{2n}{n}z^n\right\}$$

and then

$$b_n = \sum_{k=0}^{n-1} 4^{n-k-1} \frac{-1}{2k-1}\binom{2k}{k}.$$

Since (2) is true, we have

$$-2^{2n-2} \sum_{k=0}^{n-1} \frac{2^{-2k}}{2k-1}\binom{2k}{k} = \binom{2n-2}{n-1}$$

or

$$-2^{2n} \sum_{k=0}^{n} \frac{1}{2^{2k}(2k-1)}\binom{2k}{k} = \binom{2n}{n},$$

and from this the formula of Consequence 1 follows. □

**Consequence 2.** If $n \geq 1$ then

$$\sum_{k=1}^{n} \frac{1}{k}\binom{2k-2}{k-1}\binom{2n-2k}{n-k} = \frac{1}{2}\binom{2n}{n}.$$

**Proof.** If we substitute the values of $a_n$ and $b_n$ into (11), we have

$$\binom{2n-2}{n-1} = 2 \sum_{k=1}^{n-1} \frac{1}{k}\binom{2k-2}{k-1}\binom{2n-2k-2}{n-k-1}$$

or

$$2 \sum_{k=1}^{n} \frac{1}{k} \binom{2k-2}{k-1} \binom{2n-2k}{n-k} = \binom{2n}{n}. \quad \square$$

**Consequence 3.** If $n \geq 1$, then

$$\sum_{k=1}^{[\frac{n+1}{2}]} k \binom{n}{2k-1} \binom{2k}{k} 2^{n-2k} = n \binom{2n-2}{n-1}.$$

**Proof.** From

$$b_n = \sum_{k=1}^{[\frac{n+1}{2}]} k \cdot b_n^{(k)}$$

we obtain

$$\binom{2n-2}{n-1} = \sum_{k=1}^{[\frac{n+1}{2}]} k \frac{1}{n} \binom{n}{2k-1} \binom{2k}{k} 2^{n-2k}. \quad \square$$